\documentclass{mn2e}
\usepackage{psfig}
\usepackage{longtable}

\def\ltsima{$\; \buildrel < \over \sim \;$}
\def\lsim{\lower.5ex\hbox{\ltsima}}
\def\gtsima{$\; \buildrel > \over \sim \;$}
\def\gsim{\lower.5ex\hbox{\gtsima}}

\def\ltsima{$\; \buildrel < \over \sim \;$}
\def\simlt{\lower.5ex\hbox{\ltsima}} % < over ~
\def\gtsima{$\; \buildrel > \over \sim \;$}
\def\simgt{\lower.5ex\hbox{\gtsima}} % > over ~

\begin{document}

\title[Outliers]
{Are GRB 980425 and GRB 031203 real outliers or twins of GRB 060218?}

% \subtitle{}

\author[Ghisellini et al.]
{G. Ghisellini$^{1}$, G. Ghirlanda$^{1}$,
S. Mereghetti$^{2}$,
Z. Bosnjak$^{1}$,
\newauthor
F. Tavecchio$^{1}$ and
C. Firmani$^{1,3}$\\
$^1$ INAF, Osservatorio Astronomico di Brera, via E. Bianchi 46, I-23807
Merate (LC), Italy \\
$^2$ INAF, IASF--Milano, via Bassini 15, I-20133 Milano, Italy  \\ 
$^3$ Instituto de Astronom\'{\i}a, U.N.A.M., A.P. 70-264, 04510,
M\'exico, D.F., M\'exico
}
\maketitle

\begin{abstract}
GRB 980425 and GRB 031203 are apparently two outliers with respect to
the correlation  between the isotropic equivalent energy $E_{\rm iso}$ 
emitted in the prompt radiation phase and the peak frequency 
$E_{\rm peak}$ of 
the spectrum in a $\nu F_\nu$ representation (the so--called 
Amati relation).
We discuss if these two bursts are really different from the
others or if their location in the $E_{\rm iso}$--$E_{\rm peak}$
plane is the result of other effects, such as viewing them off--axis,
or through a scattering screen, or a misinterpretation of their
spectral properties.
The latter case seems particularly interesting after GRB 
060218, that, unlike GRB 031203 and GRB 980425, had a prompt 
emission detected both in hard and soft X--rays which lasted 
$\sim$2800 seconds. 
This allowed to determine its $E_{\rm peak}$ and total 
emitted energy.
Although it shares with GRB 031203 the total energetics,
it is not an outlier with respect to the Amati correlation.
We then investigate if a hard--to--soft spectral evolution 
in GRB 031203 and GRB 980425, consistent with all the
observed properties, can give rise to a time integrated 
spectrum with an $E_{\rm peak}$ consistent with the Amati relation.
\end{abstract}

\section{Introduction}

Amati et al. (2001) found a correlation between the energy emitted
(assuming isotropy) during the prompt phase of Gamma--Ray Bursts
(GRBs), $E_{\rm iso}$, and the frequency where most of this energy is
emitted, $E_{\rm peak}$ (the so called Amati relation).  Later,
Ghirlanda, Ghisellini \& Lazzati (2004) found a much tighter correlation between
$E_{\rm peak}$ and the collimation corrected energy $E_{\gamma}$, for
those bursts of known jet semiaperture angle $\theta_{\rm j}$ (the so
called Ghirlanda relation).  
These authors also showed that
all but two bursts of known redshift and $E_{\rm peak}$ are consistent
with the Amati relation.  This was confirmed more recently by
Ghirlanda et al. (2005, see also Bosnjak et al. 2006) 
with a large sample of BATSE bursts with pseudo--redshifts
(taken from Band, Norris \& Bonnell 2004), despite some
claims of the opposite (Nakar \& Piran 2005; Band \& Preece 2005).
The two {\it outliers} with respect to the Amati relation (and also
with respect to the Ghirlanda relation) are GRB 980425 and GRB 031203.
Both of them are associated with an observed supernova of type Ic,
i.e. GRB 980425--SN~1998bw (Galama et al. 1998) and
GRB 031203--SN~2003lw (Malesani et al. 2004, Thomsen et al. 2004),
but also GRB 030329 (i.e. Stanek et al. 2003), GRB 021211 (Della Valle
et al. 2002) and the very recent GRB 060218 (Campana et al. 2006; Pian
et al. 2006) are associated with a SN. 
Unlike GRB 980425 and
GRB 031203, the other bursts so far associated with a supernova event
obey the Amati relation.  
The two outliers are very close to Earth, with a
redshift of $z=0.0085$ (GRB 980425, Tinney et al. 1998) and $z=0.106$ 
(GRB 031203, Prochaska et al. 2004).  
Interestingly, GRB 060218 has an intermediate
redshift $z=0.033$ (Mirabal \& Halpern 2006; Masetti et. al. 2006),
therefore it is a factor $\sim$3 more distant than GRB 980425 and a
factor $\sim$3 closer than GRB 031203. 
Note that the other two GRBs associated with a supernova have $z=0.168$ 
(GRB 030329, Greiner et al. 2003) and $z=1.01$ (GRB 021211, 
Vreeswijk et al., 2002).

The aim of this paper is to discuss if these two bursts
are real or only $apparent$ outliers.
We will explore three different possibilities:
\begin{enumerate}
\item these two GRBs are seen off axis, as proposed 
by Ramirez--Ruiz et al. (2005);
\item what we see is the radiation surviving after having
crossed a Thomson scattering screen;
\item the real prompt emission of these bursts is subject to a
relatively strong hard to soft spectral evolution, and what we 
have seen is only the hardest, short--duration part of a
much longer emission.
\end{enumerate}

Possibility i) is quite popular, and postulates that the two outliers
are normal GRBs observed off axis, at a viewing angle $\theta_{\rm v}$
of the order of twice the aperture angle of the jet $\theta_{\rm j}$.
(Ramirez--Ruiz et al. 2005; see also other off--axis models by
Yamazaki, Ionetoku \& Nakamura, 2004; Eichler \& Levinson 2004) The
appeal of this idea is that, within the homogeneous jet scenario, such
outliers {\it must} exist, and should outnumber ``normal" GRBs by a
large factor, given by the solid angle ratios.  Of course, since the
received radiation for off--axis observers is strongly dimmed by
de--beaming, the actual number of observable outliers depends on an
accurate calculation of how both the peak energy and the total time
integrated flux depend on the viewing angle $\theta_{\rm v}$, the jet
aperture $\theta_{\rm j}$ and the bulk Lorentz factor $\Gamma$.

Possibility ii) postulates the existence of some scattering material
along the line of sight, following the original idea of
Brainerd (1994) and Brainerd et al. (1998), and later used
also by Barbiellini et al. (2004).
In this case we are receiving the transmitted flux piercing through
the scattering screen.  The received flux can be much dimmer than
the incident one, and the energy decline of the (Klein--Nishina)
scattering cross section imprints an important modification on the
transmitted spectrum making it harder and peaking at a larger 
$E_{\rm peak}$.  
The intrinsic spectrum of the two outliers is then softer and
brighter, and thus possibly consistent with the Amati relation.

Case iii) is based on spectral evolution.
For GRB 031203, in fact, there is a strong observational
evidence that what INTEGRAL has seen is only a part of
a more complex story: besides the high energy emission
above 20 keV, observed to a have a flat spectrum and lasting
for $\sim$30 seconds, there must be a significant emission
at much softer energies, lasting for much longer 
(i.e. more than a few hundred seconds, see section 4). 
This has been inferred through the X--ray flux 
scattered by dust layers in our Galaxy and producing
time variable expanding rings around the source (Vaughan et al. 2004).
The fluence of this component may have exceeded the fluence
above 20 keV, and this supports the idea that it belongs to
the prompt emission.
The peak frequency of the combined soft and hard components,
albeit uncertain, could well be in the range 3--10 keV,
making GRB 031203 consistent with the Amati relation.
In exploring this idea, we find strong similarities
(and some differences) with the behaviour of a recent
bursts, observed by SWIFT, namely GRB 060218.
Its prompt emission lasted for $\sim$3000 seconds, enabling
the XRT instruments to follow most of it: for this burst we then 
have direct information of the high and low energy spectral
components. 
Spectral fitting of the combined (XRT+BAT) spectra yields a peak 
frequency perfectly consistent with the Amati relation.
Since its bolometric isotropic emitted energy is almost the
same of GRB 031203, it is natural to consider them as twins.

Then the question: is it possible that even GRB 980425 
had a similar behaviour? 
Is it possible that the prompt emission
as observed by BATSE on one hand, and by the WFC and GRBM
instruments of $Beppo$SAX on the other, was only
part of a more complex story, and that the time
integrated $E_{\rm peak}$ is much smaller than what
measured in the first $\sim$50 seconds?

To determine if the two outliers are really so or they only
{\it appear} as such, is important for several reasons.

For instance, Paczynski \& Haensel (2005) recently proposed that all
``classical" GRBs are associated to the transition from a neutron to a
quark star.  The two outliers, in this framework, should correspond to
massive stars outgoing the SN Ic explosion {\it without} undergoing
the transition to the quark star.  Clearly, in this view, GRB 980425
and GRB 031203 are real outliers, emitting their prompt radiation with
different mechanisms than the other, ``classical" GRBs. 
Moreover,
if these two bursts are real outliers they should have a flatter
luminosity function with respect to ``normal'' GRBs in order to
account for their lower detection rate, as recently discussed by
Liang, Zhang and Dai (2006).

On the contrary, if the two GRBs are not outliers, than we
strengthen the Amati relation, which becomes more universal.
Its physical interpretation therefore becomes even more 
compelling.

\section{Homogeneous jet seen off axis}

\begin{figure}
\vskip -0.5 true cm
\psfig{figure=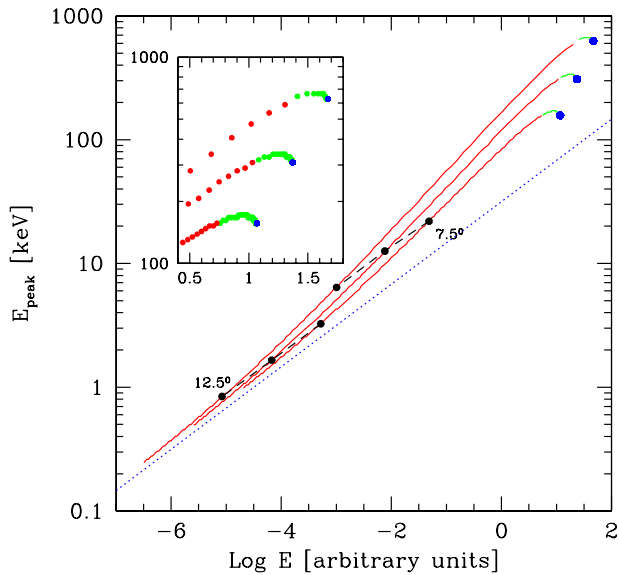,angle=0,width=9cm}
\vskip -0.5 true cm
\caption{
The peak of the observed spectrum $E_{\rm peak}$ as a
function of the time integrated flux $E$. 
Both depend on the viewing angle $\theta_{\rm v}$.  In the
insert we show a zoom for small viewing angles, within (green dots)
and outside $\theta_{\rm j}$.  We assumed $\theta_{\rm j}=5^\circ$,
$0<\theta_{\rm v}<20^\circ$, $\alpha_1=0.5$, $\alpha_2=2$.  The dotted
line, shown for comparison, has a $1/3$ slope.  The three lines have
$\Gamma=50$, 100 and 200, and $E^\prime_{\rm peak}=1.25$ keV. 
Black points, connected with dashed lines, correspond to the same
viewing angle $\theta_{\rm v}=(7.5,12.5)$ for the three different
choices of $\Gamma$.
} 
\label{spectrum}
\end{figure}

Assume that the radiation is collimated into a cone of semiaperture
angle $\theta_{\rm j}$ and that the angle between the jet axis
and the line of sight is $\theta_{\rm v}$.  
The fireball emits, in its comoving frame, a time integrated spectrum 
described by a double power law,
smoothly joining at some frequency $\nu^\prime_{\rm c}$. 
This function is a simplified version of the popular Band model (Band
et al. 1993) composed by two powerlaws smoothly joined by an
exponential cutoff. 
Calling $x^\prime \equiv \nu^\prime/\nu^\prime_{\rm c}$, the time integrated 
energy spectrum, ${\cal E}^\prime(x^\prime)$, is
\begin{equation}
{\cal E}^\prime(x^\prime) \, =\, k\, { {x^\prime}^{-\alpha_1} 
\over 1 +{x^\prime}^{\alpha_2-\alpha_1}}
\end{equation}
where $\alpha_1$ and $\alpha_2$ represent the power laws' energy
indices below and above $\nu^\prime_{\rm c}$, respectively. 
Since the velocity vectors of the fireball are radially distributed 
within $\theta_{\rm j}$, we need to integrate, with the appropriate 
beaming factor, over the entire fireball surface in order to 
derive the energy spectrum observed at $\theta_{\rm v}$.  
Calling $\delta\equiv [\Gamma (1-\beta\cos\theta)]^{-1}$ we have
\begin{equation}
E(x, \theta_{\rm v}) \, =\, 
\int^{\theta{\rm v}+\theta_{\rm j}}_{\max(0,\, 
\theta{\rm v}-\theta_{\rm j})}
\Delta\psi \,  \delta^2
{\cal E}^\prime(x/\delta)\sin\theta d\theta 
\end{equation}
Within the limits of integration, 
the factor $\Delta\psi$ is given by (see Ghisellini \& Lazzati 1999)
\begin{eqnarray}
\Delta \psi \, &=&\, 2\pi; {\hskip 4.45 true cm}
\theta<\theta_{\rm j}-\theta_{\rm v}\nonumber \\
\Delta \psi \, &=&\, \pi+2\arcsin \left({\theta_{\rm j}^2-
\theta_{\rm v}^2- \theta^2 \over 2\theta \theta_{\rm v}}\right); 
\quad \theta\ge \theta_{\rm j}-\theta_{\rm v}
\end{eqnarray}
We then define $E_{\rm peak}(\theta_{\rm v})$ as the maximum of the
$xE(x,\theta_{\rm v})$ function.  In Fig. \ref{spectrum} we plot
$E_{\rm peak}$ as a function of the time integrated flux
$E(\theta_{\rm v})=\int E(x,\theta_{\rm v}) dx$, having assumed
$\theta_{\rm j}=5^\circ$, and varying $\theta_{\rm v}$ between
$0^\circ$ and $20^\circ$.  The three curves correspond to $\Gamma=50$,
100 and 200.  It can be seen that the $E_{\rm peak}\propto E^{1/3}$
behaviour (dotted line in Fig. \ref{spectrum}) is reached only for
large viewing angles.  In the insert, the zoom for small viewing
angles shows that $E$ changes by a factor 2 for 
$0<\theta_{\rm v}<\theta_{\rm j}$, while $E_{\rm peak}$ 
remains approximately constant.

The reason for the departure from the $E_{\rm peak}\propto E^{1/3}$
behaviour is the following 
(see also Eichler \& Levinson 2004 and Toma et al. 2005):
for large viewing angles, there is a small
difference between the flux received from the near and the far edges
of the jet, which therefore contribute equally to the total flux.  As
long as this is the case, we have $E_{\rm peak}\propto E^{1/3}$.  As
the viewing angle is decreased, the difference of the fluxes received
from the edges increases, with the far edge becoming progressively
negligible (with respect to the near edge), causing a deficit of total
flux with respect to the extrapolation from larger viewing angles.

The reason of the factor 2 difference between the energy
observed at $\theta_{\rm v}=0$ and $\theta_{\rm v}=\theta_{\rm j}$
is the following: on axis observers receive most of the flux
from a circle of angular radius $1/\Gamma$.
At the border of the jet (i.e. for $\theta_{\rm v}=\theta_{\rm j}$)
only $\sim$half of the circle is inside the jet and emitting.
The maximum value of $E_{\rm peak}$ is not observed by 
observers perfectly on--axis, but slightly off.
To explain this effect consider the on--axis observer: 
in this case the maximum flux is again received by the ring  
of angular radius $1/\Gamma$, with contributions
both from smaller angles (with larger $E_{\rm peak}$)
and larger angles (with smaller $E_{\rm peak}$).
The observed $E_{\rm peak}$ is found by
integrating the spectra over the jet solid angle.
For slightly off--axis observers, the contribution
from large angles (on one side of the jet) is partially missing 
(lying otside the jet solid angle), while it appears a contributions
from still larger angles from the opposite side of the jet.
This latter component is too de--beamed to compensate for the missing one
and the net effect is to slightly increase
the observed $E_{\rm peak}$.

\begin{figure}
\vskip -0.5 true cm
\psfig{file=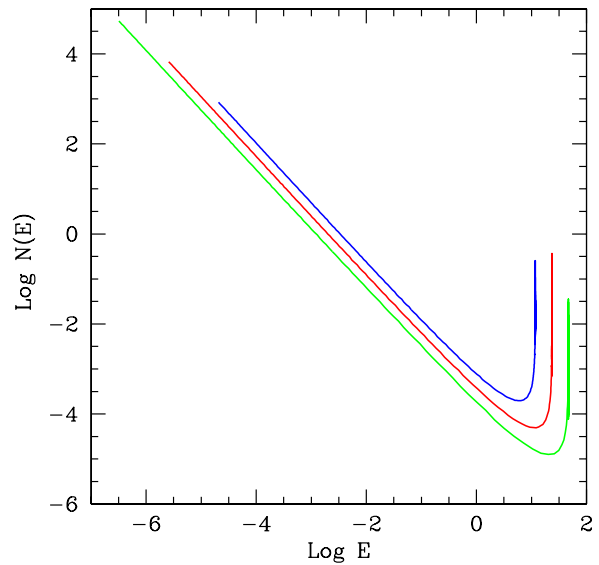,width=9truecm}
\vskip -1 true cm
\caption{
The ``energy function" $N(E)$ (i.e. the time integrated
luminosity function) for the 3 cases corresponding to
$\Gamma=$50, 100 and 200 (upper, intermediate and lower lines) shown
in Fig. \ref{spectrum}.  
}
\label{pl}
\end{figure}

\subsection{The ``energy" function}

Assume that a subset of bursts have all the same 
radiated energy $E_{\rm on}$ if seen on axis.
Bursts observed off--axis will be observed to emit an energy $E$,
and their number is proportional to the corresponding solid angle
(see Urry \& Shafer 1984 and Celotti et al. 1993 for blazars):
\begin{equation}
N(E, E_{\rm on}) dE \, =\, {d\Omega \over 2\pi}\,  =\, \sin\theta d\theta
\end{equation}
leading to 
\begin{equation}
N(E, E_{\rm on} ) \, =\, \left( {dE \over d\cos\theta} \right)^{-1}
\end{equation}
If bursts have an intrinsic
distribution $\Phi(E_{\rm on})$ of energetics as seen on axis, than we
have to integrate $N(E, E_{\rm on})$ over that distribution.

Fig. \ref{pl} shows $N(E,E_{\rm on})$ for the three cases shown 
in Fig. \ref{spectrum} (i.e. $\Gamma=50$, 100 and 200).
The resulting function is a power law at small energies,
and note the increase towards the upper end of the
energy range.
This corresponds to observers within the jet aperture angle,
who see almost the same energetics.
At small energies, the slope of the power law is $4/3$.
This value has a simple explanation. 
For large viewing angles, in fact, the observed energy
$E\propto \delta^3\propto [\Gamma (1-\cos\theta)]^{-3}$.
Therefore $(dE/d\cos\theta)^{-1}\propto \delta^{-4} \propto E^{-4/3}$,
which is a very good approximation of the numerical results
shown in 
Fig. \ref{pl}\footnote{
This results is general, as demonstrated by Urry \& Shaefer
(1984): for beaming amplification factors $L\propto \delta^p$, the
resulting low luminosity branch of the observed luminosity function
$N(L) \propto L^{-1-1/p}$.}.

\subsection{Application to GRB 980425 and GRB 031203}

If GRB 980425 and GRB 031203 were off--axis events and
we require their on--axis energetics to satisfy the Amati
relation, then we end up to isotropic energies larger
than $10^{53}$ erg (even larger than suggested by Ramirez--Ruiz
et al. 2004 for GRB 031203 using a 
$E_{\rm iso} \propto E_{\rm peak}^{1/3}$
approximation).

All models invoking an intrinsic very large energy output, similar or
even larger than the energy output of ``classical" GRBs, suffer from a
severe problem: the two outliers are the closest of all GRBs and it is
therefore extremely unlikely that such energetic bursts exist in our
vicinity, for all reasonable luminosity functions (or energy
functions).  This issue has been discussed in Guetta et al. (2004) and
with the detailed ``trajectories" calculated in the previous sections
the problem is even worse than assuming the naive $E_{\rm peak}\propto
E_{\rm iso}^{1/3}$ behaviour, since the steeper dependence requires
larger values of $E_{\rm iso}$ to become consistent with the Amati
relation.  For this reason, we discard this possibility.

On the other side, if jets are homogeneous, we {\it must} observe some
of them slightly off axis.  For a given sensitivity threshold, the
amount of them depends on the specific value of the bulk Lorentz
factor $\Gamma$, since larger $\Gamma$ make the received flux to
decrease more rapidly increasing the viewing angle.  This can be seen
in Fig. \ref{spectrum}, where we have marked the points corresponding
to the same viewing angles: increasing by a factor 2 the bulk Lorentz
factor implies to decrease by an order of magnitude the measured
$E_{\rm iso}$.  Fig. \ref{pl} makes this argument more quantitative,
showing the relative number of expected off--axis sources with respect
to on--axis ones.

This is not meant to discard the fact that some GRBs can
be seen off--axis, but only to discard the fact that this is
the only reason for bringing the two outliers over the
Amati relation.

\section{Piercing through a scattering screen}

As already discussed, the vicinity of the two outliers 
argues against these two bursts being normal powerful bursts
 seen off--axis, 
(with, say, $E_{\rm iso}\sim 10^{53}$ erg
or more, as suggested by Ramirez--Ruiz et al. 2004)
since any reasonable luminosity function 
would make them extremely rare in the volume corresponding
to their redshifts.
Instead, their vicinity argues in favour of an intrinsically small
isotropic energy and luminosity (see e.g. Guetta et al. 2004).
We then investigate wether these two bursts are observed through a 
screen of large 
optical depth, which at the same time decreases their apparent 
isotropic energy and {\it increases} their apparent $E_{\rm peak}$.
Their intrinsic emission (i.e. before passing
through the scattering screen) could then be consistent with 
the Amati relation.

\subsection{Transmitted spectra}

The spectrum, incident on a screen of large optical depth, is
assumed to be a smoothly joined broken power law (with energy indices
$\alpha$ and $\beta$ at low and high frequencies, respectively)
as in Sec. 2, or a single power law ending with a an exponential cut:
\begin{eqnarray}
F(x) \, &=&\, F_0 \, { (x/x_b)^{-\alpha} \over
1+ (x/x_b)^{\beta-\alpha} }; 
\nonumber \\
F(x) \, &=&\, F_0 \, x^{-\alpha} \exp(-x/x_c)
\end{eqnarray}
where $x\equiv h\nu/(m_{\rm e} c^2)$ is the photon energy
in units of the electron rest-mass energy.
We assume that $\alpha<1$ and $\beta>1$.
Thus, in a $xF(x)$ representation, the peak energy $x_{\rm peak}$ is
$x_{\rm peak} = x_b\, [(1-\alpha)/(\beta-1)]^{1/(\beta-\alpha)}
$
for the broken power law case and
$x_{\rm peak}= x_c (1-\alpha)$ for the cut--off power law case.

Along the line of sight the Thomson optical depth is
$\tau_{\rm T}=\sigma_{\rm T} n \Delta R$, where $n$ is the number 
density and $\Delta R$ is the typical size of the scattering region.  
We will
consider values of $\tau_{\rm T}$ even larger than one, and this is possible
only if the scattering region is close to the burst, and becomes
almost completely ionized by the prompt emission.  
Otherwise, we would
detect very large column densities in the X--ray spectra of the
afterglow.  The Klein Nishina optical depth is $\tau_{\rm KN} =
\tau_{\rm T}\sigma_{\rm KN}/\sigma_{\rm T}$.  
The {\it transmitted}
luminosity is the luminosity passing through the scattering material
without suffering any scattering:
\begin{equation}
F_{\rm t}(x) \, =\, F(x) \exp(-\tau_{\rm KN})
\end{equation}
\begin{figure}
\vskip -0.5 true cm
\centerline{\psfig{figure=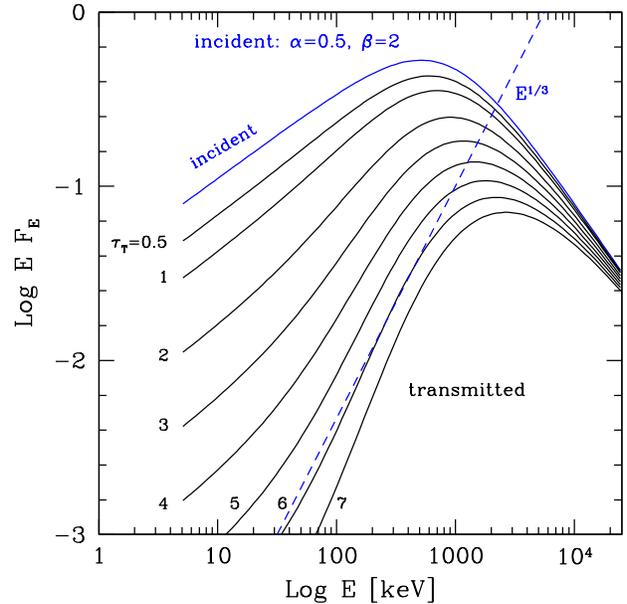,angle=0,width=9.5cm}}
\vskip -0.5 true cm
\caption{Transmitted spectrum for different values of the Thomson
optical depth $\tau_{\rm T}$, as labelled.  The incident spectrum has
$E_{\rm peak}=511$ keV, $\alpha=0.5$ and $\beta=2$.  The chosen value
of $\alpha$ corresponds to the hardest synchrotron slope produced by a
{\it cooling} population of electrons.  The dashed line, $\propto
E^{1/3}$ corresponds to the hardest possible synchrotron slope
produced by an electron distribution with a low energy cut--off, and
no cooling.  
Harder spectra are difficult to reconcile with the standard synchrotron
interpretation (Ghirlanda, Celotti \& Ghisellini 2003).  Note that we can
have transmitted spectra harder than $\alpha=-0.3$, even with incident
spectra having $\alpha=0.5$, for $\tau_{\rm T}>6$.  }
\label{transmitted}
\end{figure}
\begin{figure}
\vskip -0.3 true cm
\centerline{\psfig{figure=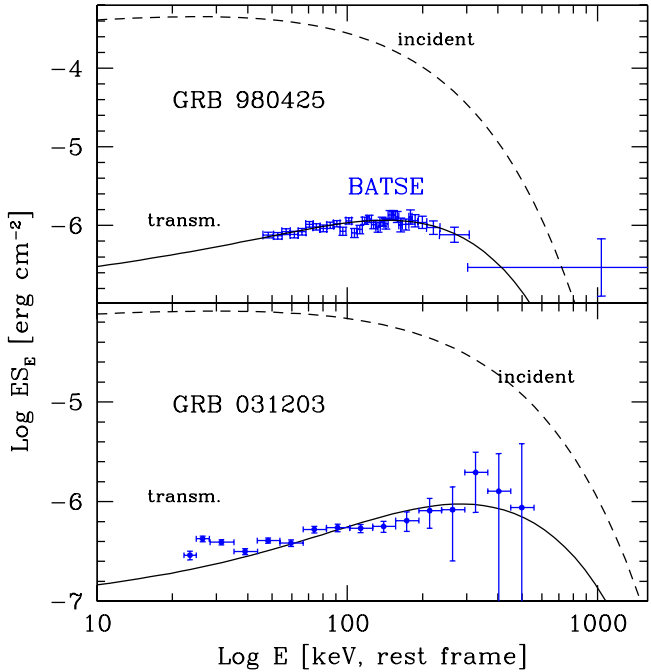,angle=0,width=10cm}}
\vskip -0.5 true cm
\caption{ 
The observed $ES(E)$ spectra of the prompt emission of GRB
980425 (top panel) as observed by BATSE and of GRB 031203 (bottom
panel) as observed by INTEGRAL, expressed in fluence units.  The data
shown for GRB 980425 have been analyzed by us using a power law with
an exponential cut--off to fit the time integrated spectrum over
the $\sim 30$ sec duration of the burst.  For GRB 031203, we report
the spectrum analyzed by us, using version 5.1 of the OSA
analysis software and the corresponding response matrices. 
This spectrum corresponds to the first 20 sec of the emission as the
following 20 to 40 sec do not contribute significantly.  
Solid lines corresponds to the flux passing through a screen of material, 
while dashed lines are the intrinsic flux (i.e. before passing through 
the screen).  
The parameters of the model are listed in Table 1.  
}
\label{outliers}
\end{figure}
\begin{figure}
\vskip -0.5 true cm
\centerline{\psfig{figure=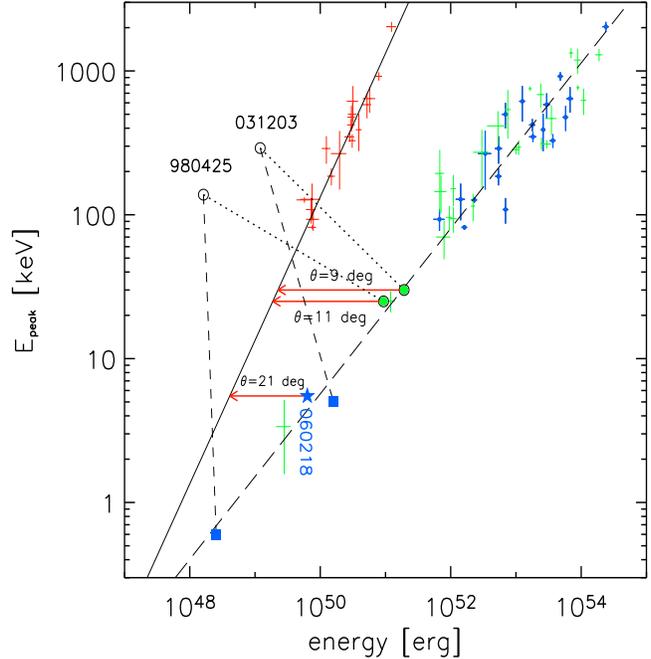,angle=0,width=9.5cm,height=9.7cm}}
\vskip -0.3 true cm
\caption{ The correlation between $E_{\rm peak}$ and the isotropic
equivalent energy radiated during the prompt emission $E_{\rm iso}$
(blue and green symbols, the ``Amati" relation, Amati et al. 2002))
and the collimation corrected energy 
$E_\gamma=(1-\cos\theta_{\rm j})E_{\rm iso}$, 
where $\theta_{\rm j}$ is the semiaperture angle of
the jet, assumed conical (red crosses, the ``Ghirlanda" relation,
Ghirlanda et al. 2004, but here shown in the case of a circumburst
density profile $n\propto r^{-2}$ appropriate for a stellar wind made
by the progenitor, see Nava et al. 2006).  Blue points are the 19
GRBs with a firm estimate of the jet opening angle; green points are
those bursts (adapted from Amati et al. 2006) without a firm estimate
of $\theta_{\rm j}$.
GRB 980425 and GRB 031203 are outliers (open circles) 
with respect to the ``Amati" relation (long--dashed line), 
but they can be made consistent with this correlation 
dotted line and green filled circles) by assuming that the radiation
we see is the transmitted radiation crossing a scattering screen as
illustrated in Fig.~\ref{outliers}.  
Without the scattering screen,
these two bursts would have appeared as intermediate between X--ray
rich GRB and X--ray flashes.  Moreover, they can be made consistent
with the Ghirlanda relation ( solid line, in the case of a
``wind" density profile) if their jet opening angle is $9^\circ$ and
11$^\circ$, as labelled.  GRB 060218 would be consistent with the
Ghirlanda relation for $\theta_{\rm j}\sim 21^\circ$. 
Alternatively we also show the effect of spectral evolution (Sec. 4)
which can make the two outliers consistent with the ``Amati'' relation
(short--dashed line and blue--filled squares) if we only detected a
short portion of their much longer emission which rapidly evolved from
the hard $\gamma$--ray band to the soft X--ray band. 
In this case, if
we require that these two bursts are consistent with the ``Ghirlanda''
relation we derive angles of 34$^\circ$ and 13$^\circ$ for GRB 980425
and GRB 031203, respectively. 
}
\label{correla}
\end{figure}

Fig. \ref{transmitted} shows the transmitted spectra 
for different values of $\tau_{\rm T}$ for an 
incident spectrum peaking at $E_{\rm peak}=511$ keV,
with $\alpha=0.5$ and $\beta=2$.
Incidentally, note that the assumed $\alpha$ corresponds 
to the spectrum produced by a cooling population of electrons,
but with an initial low energy cutoff
(see Ghisellini, Celotti \& Lazzati 2000).
The effect of scattering 
hardens the transmitted spectrum, to reach values
harder than $\alpha=-1/3$ (the ``line of death"
for the synchrotron interpretation, Preece et al. 1998),
for $\tau_{\rm T}>6$.

The other effect imprinted on the transmitted spectrum by the
scattering screen is the shift of the peak energy $E_{\rm peak}$ to
higher values, as shown in Fig. \ref{transmitted}. 
This is due to the declining (with energy)
Klein--Nishina cross section $\sigma_{\rm KN}$. 
For instance, the transmitted spectrum in the case of $\tau_{\rm T}=6$
has $E_{\rm peak}\sim 2$ MeV, while the intrinsic spectrum has 
$E_{\rm peak}= 511$ keV.

Note also that the bolometric transmitted energy,
for this particular case of an intrinsic $E_{\rm peak}=511$ keV,
decreases much less than the factor $exp(-\tau_{\rm T})$
expected in the Thomson regime (a factor $\sim$10 for $\tau_{\rm T}=5$,
instead of the factor 150 we would have in the Thomson regime).

\subsection{Application to GRB 980425 and GRB 031203}

In Fig. \ref{outliers} we show the observed spectrum
of the two sources (time integrated over the duration of
the bursts), the assumed intrinsic spectrum, and the
transmitted one.

We reanalyzed the INTEGRAL data obtained with the IBIS/ISGRI
instrument using version 5.1 of the OSA analysis software and
the corresponding response matrices. 
The spectrum corresponding to the first 20 seconds of the burst
(start time 2003--12--03 22:01:26) can be fit by a power law
with photon index 1.7 and fluence 1.4e-6 erg cm$^{-2}$.

The found slope is consistent with the slope found by Sazonov et
al. (2004), but smaller by a factor 1.4 in normalisation.

For GRB 980425 we have analysed the BATSE data (to show the data
points) using a cut--off power law model. 
The best fit parameters are
$\alpha=0.45\pm0.22$ and $E_{\rm peak}\sim 138$ keV 
($\chi^2=109/97~$dof), giving a flux of
$\sim 10^{-7}$ erg/cm$^2$ sec (40--700 keV), 
consistent with that reported in Pian et al. 2000.

The same model (power law + exponential cut) was assumed for GRB
031203, for which we plot the spectrum analyzed by us in
Fig. \ref{outliers}, but with frequencies in the source rest frame.  
For GRB 031203 a
soft component has been inferred through the observed transient rings
produced by the scattering of the burst radiation by dust layers
in our own Galaxy (Vaughan et al. 2004, Tiengo \& Mereghetti 2006 -- see
also Vaughan et al. 2006 for another case of dust scattering halo
associated with GRB 050724).  
This component, that will be discussed in the next 
section, is here neglected. 
\begin{table}  
\begin{center}  
\begin{tabular}{|l|lllllll|}  
\hline      
GRB & $\tau_{\rm T}$ & $\alpha$   & $E_{\rm c}$  
& $E^{\rm intr}_{\rm peak}$  & $E_{\rm peak}^{\rm obs}$  
& $\theta_{\rm j}$ &$f$ \\
&  & &  &keV &keV & & \\  
\hline  
% 980425    &6.8  &0.6    &85   &34  &138 &$18^\circ$ &292 \\
980425     &7.5  &0.7    &83.3 &25  &138 &$11^\circ$ &577 \\
031203     &6.5  &0.85   &200  &30  &292 &$9^\circ$  &161 \\
% 031203    &5    &0.6    &175  &70  &242 &$16^\circ$   &45 \\
% 031203    &5    &0.8    &300  &60  &351 &$16^\circ$?? &41 \\
% 031203    &6    &0.7    &183  &55  &265 &$N^\circ$    &82 \\
%
\hline  
\end{tabular}  
\caption{Input parameters of the scattering model for the two
outliers.  The parameter $f$ is the ratio of the intrinsic to
transmitted total energy.  The jet opening angle $\theta_{\rm j}$ is
derived requiring that the the intrinsic $E_{\rm peak}$ and $E_\gamma$
fit the ``Ghirlanda" relation derived under the wind circum--burst
density hypothesis (Nava et al. 2006).
} 
\end{center}  
\end{table}  

It can be seen in Fig. \ref{outliers} that the model can reproduce the
observed data quite satisfactorily.  In Tab. 1 we report the assumed
parameters ($\tau_{\rm T}$, $\alpha$, $E_{\rm c}$) for GRB 980425 and
GRB 031203, and in Fig. \ref{correla} we show  where these two
bursts ``move" in the $E_{\rm peak}$--$E_{\rm iso}$ plane (filled circles)
when
the incident spectrum (i.e. that impinging on the scattering screen)
is reconstructed with the assumed parameters. 
In Fig. \ref{correla} we also show what jet opening angle $\theta_{\rm j}$ 
they should have to also fit the ``Ghirlanda" relation between
$E_{\rm peak}$ and the collimation corrected energy
$E_\gamma=(1-\cos\theta_{\rm j}) E_{\rm iso}$,  as derived under
the hypothesis of a wind density environment (Nava et al. 2006).
This angle turns out to be $\sim 9^\circ$ for GRB 031203 and 
$\sim 11^\circ$ for GRB 980425. 
These values of $\theta_{\rm j}$ lie in
the large--angle tail of the distribution of known jet opening angles
(for 19 GRBs - see Fig. 6 in Nava et al 2006). In fact, the low peak
energies of the intrinsic spectra (i.e. before piercing through the
scattering screen) places both GRB 980425 and GRB 031203 in the XRF
region of the $E_{\rm peak}$--$E$ plane of Fig. \ref{correla} where the
``Amati'' and ``Ghirlanda'' correlation converge. 
Bursts in this
region have, on average, jet opening angles larger than those
placed at the opposite extreme (such as e.g. GRB 990123,
the most energetic burst in the $E_{\rm peak}$--$E_{\rm iso}$
and in the $E_{\rm peak}$--$E_\gamma$ planes).

Although this ``screen model" might have interesting feature, 
it suffers from some problems, related to the presence of an
optically thick screen itself.
The first problem is why this material is present in some
bursts but not in others.
An additional problem concerns its location:
if it is at some distance from the bursts, its material
would not be ionized (and the dust in it not destroyed)
by the prompt radiation, making the optical afterglow 
not detectable.
If it is instead located close to the burst, one should see
some signature of its presence when the fireball and/or
the supernova ejecta collide with it.

\section{Spectral evolution}

The scenarios presented in the previous sections were
aimed to explain the prompt emission as observed by
BATSE and $Beppo$SAX for GRB 980425, and by
INTEGRAL for GRB 031203, assuming that the received
prompt emission was mainly in the 20--300 keV range.
This may not be true, however, for the following reasons:

\begin{itemize}
\item 
The echo--rings detected by XMM--Newton a few hours after the trigger
of GRB 031203 can be produced only if the incident flux is at soft
X--ray energies and large.  
Reconstructing the flux, fluence and
spectrum of the emission which was scattered by the dust layers of
our Galaxy is not straightforward, but the existing estimates point
towards a total fluence at least as large as the fluence detected by
INTEGRAL, at energies around 1 keV (Watson et al. 2006; Tiengo \&
Mereghetti 2006).  In this respect, note that the re--analysis of the
INTEGRAL data performed by us, with improved detector response
matrices yields a total fluence a factor 1.4 less than what quoted by
Sazonov et al. (2004), while the spectral shape is the same.

\item
One important issue is if this soft component belongs to the
prompt emission or if it is instead the beginning of the afterglow.
The large fluence, however, suggests that this is part of the
prompt, since we never observed X--ray afterglows as bright as the
prompt in any other GRB.

\item
Another issue concerns the duration of the soft component. 
A lower limit can be roughly estimated based on  the
the lack of detection with IBIS/ISGRI 
in the images obtained after (and before) the
GRB prompt emission. Such images are limited
to the energy range above $\sim$17 keV. 
Therefore one has to invoke either that the 
soft X--ray emission desumed from the dust 
scattering rings had a spectrum with a very
sharp cut--off before the IBIS/ISGRI energy
range or that the soft X--ray emission lasted
more than several hundred seconds, in order to 
give the required fluence without violating
the flux upper limits. 
Note that a duration of the order of a few thousand 
seconds would still be compatible with the observed 
width of the expanding rings measured with XMM--Newton.

\item
For GRB 980425 we do not have direct evidences of 
the presence of a (possibly long--lasting) soft component,
but we do have some indication of spectral softening from the
analysis of the WFC (Wide Field Camera) and GRBM monitor
on board $Beppo$SAX, as presented by Frontera et al. (2000),
see the spectra reported in Fig. \ref{980425}.
The total fluence in the WFC [2--28 keV] is of the same
order than the fluence in GRBM (which agrees with BATSE). 

\item
The idea of an hard--to--soft spectral evolution, with a very long
soft emission, received a strong support from the recently {\it Swift}
GRB 060218 (Cusumano et al. 2006).  For this burst the narrow field
instruments of Swift could follow the emission for most of the prompt
phase, which lasted at least $\sim$3000 seconds (Barthelmy et
al. 2006).  This burst showed a hard--to--soft spectral evolution, and
its time integrated spectrum has a peak energy around 5 keV and a
$\sim 6\times10^{49}$ erg isotropic energy which makes this burst
fully consistent with the Amati relation (Campana et al. 2006;
see Fig. \ref{correla}).  
The total energetics of GRB 060218 is very similar to the one of GRB
031203, and both are associated with a supernova of similar power
(e.g. Cobb et al. 2006).  
This burst prompted us to explore in some
detail the idea of spectral evolution for the two outliers.

\item
Finally we point out that, recently, a tight correlation
was discovered (Firmani et al. 2006), 
involving only observables belonging to the
prompt emission: the peak luminosity, $E_{\rm peak}$ and
the ``high signal" timescale $T_{0.45}$ (it is the same
timescale used for the characterization of variability, see 
Reichart et al. 2001).
GRB 980425 and GRB 031203 are outliers with respect to this correlation.
However GRB 031203 would become consistent with 
this newly found correlation if $E_{\rm peak}$ were in the 
$\sim$keV range, as suggested by the echo--rings.

\end{itemize}

In this section we see if it is possible to construct a
simple model for the spectral evolution of GRB 031203 and GRB 980425
that is at the same time consistent with the existing observations of
these two GRBs and also predicting, for their time integrated spectra,
values of $E_{\rm peak}$ and $E_{\rm iso}$ in agreement with the Amati
relation.  We are inspired to do such a modelling by the behaviour of
GRB 060218, and we will then first construct our toy model in order to
reproduce the time-dependent behaviour of its spectra.  For this burst
we have much more information than for the others, and the modelling
therefore merely implies to characterise the spectral index, the
cut--off energy and the normalisation of the spectrum with smooth
functions of time, to reconstruct both the time resolved spectra and
the light curves in the XRT [0.3--10 keV] and BAT [15--150 keV] energy
bands.

For GRB 031203 the time dependent information is almost not existent:
we assume that the soft X--ray spectrum and fluence, around 1 keV, is
is the one inferred (albeit in an approximate way), by the echo--rings
observed by XMM--Newton.

For GRB 980425 we have even less information, and the modelling is
done only to show that there is no evidence against a hard--to--soft
evolution also for this burst.

We are aware that 
i) this model does not pretend to be unique, since
it can be one of many spectral evolution models; 
ii) it is completely phenomenological, and it has no theoretical support, 
iii) for the sake of simplicity, we will assume that the spectral 
indices and the high energy cut off of the spectrum, as well as its 
normalisation, are smooth functions of time (i.e. we will assume 
no flares), and finally
iv) even if we will minimise the number of free parameters, in the end
we need many.

These can be taken as strong limitations if we were really seeking
``the" model for interpreting the properties of spectral evolution,
but this is not our aim.  Our goal is instead to demonstrate that,
even with a naive and simple approach, it is indeed possible to fulfil
the two requirements of being in agreement with all observations and
nevertheless making these specific bursts to obey the Amati relation.

\subsection{The case of GRB 060218}

\begin{figure*}
\vskip -1 true cm
\centerline{\psfig{file=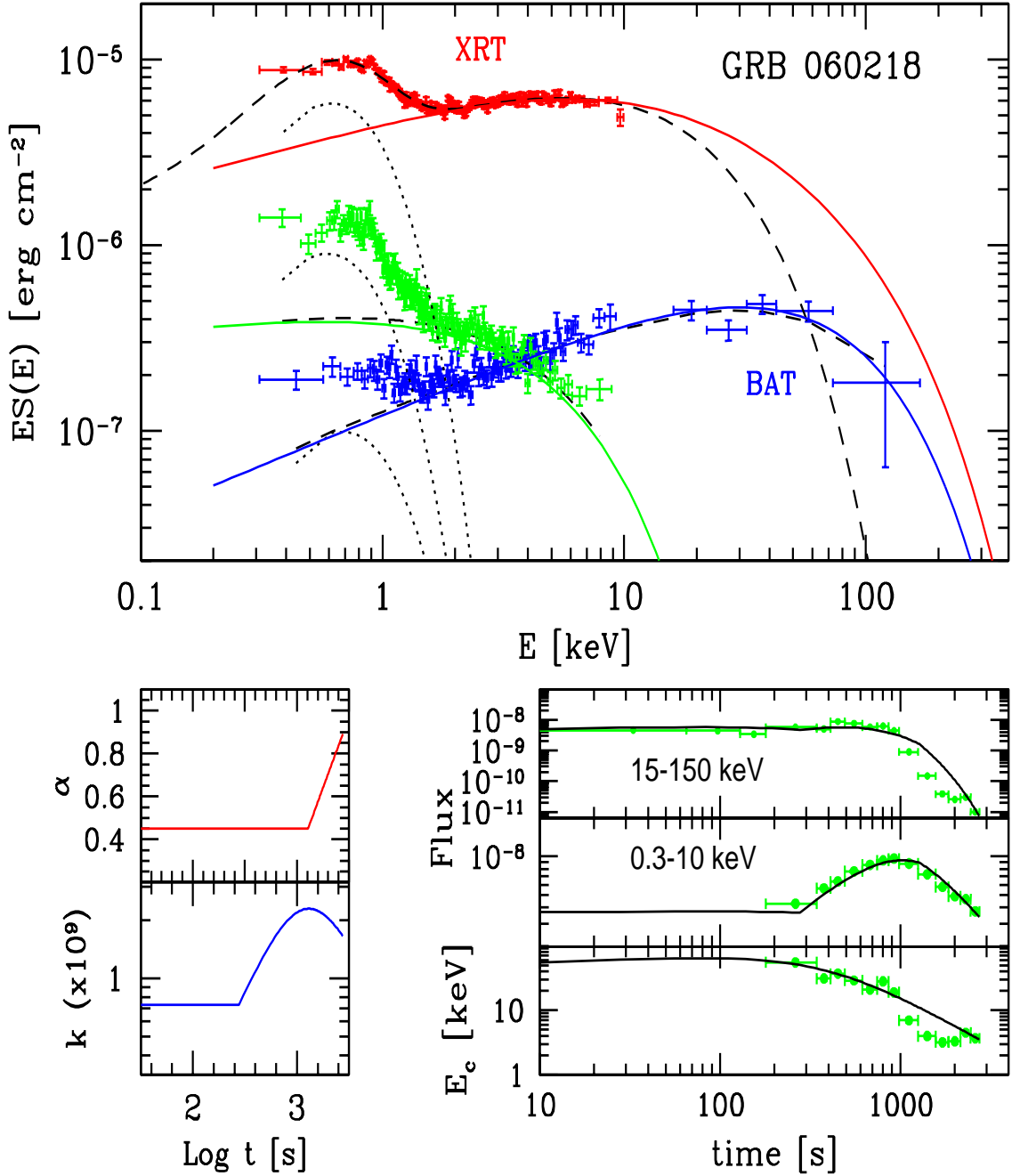,width=17truecm,height=21truecm}}
\vskip -1.5 true cm
\caption{
Top panel: Spectra of GRB 060218 for different time--bins: 
i) entire duration (top); 
ii) [159--309 s] (rising spectrum with also BAT data) 
iii) [2456--2748 s] (soft spectrum).  
We plot $ES(E)$ vs $E$, $S(E)$ being the fluence. 
Dotted lines indicate the blackbody component, not considered 
for the spectral evolution, and long--dashed lines represents 
the best fit obtained from the analysis of the data.  
Continuous lines show the results of our proposed
modelling.  
Left bottom panel: assumed behaviour of the normalisation
$K$ and energy spectral index $\alpha$.  Right bottom panel: light
curves in the BAT (15--150 keV) and XRT (0.3--10 keV) range, and
evolution of $E_{\rm c}$.  The flux in the 0.3--10 keV is the
(de--absorbed) flux of the cut--off power law component only: we have
subtracted the blackbody component from the total flux.  Continuous
lines are the results of our modelling.  
}
\label{060218}
\end{figure*}

Fig. \ref{060218} (upper panel) shows the time integrated spectrum of
GRB 060218. 
We re--analyzed the XRT data of the first $\sim$2600
sec (in $wt$ mode) starting $\sim 159$ sec after the BAT trigger
(03:34:30 UT - Cusumano et al. 2006). For the data extraction we used the
standard {\it xrtpipeline} (v.0.9.9) and for the spectral analysis
we used the v.007 detector response
matrix (for {\it wt} mode limited to grades 0--2).  
Campana et al. 2006, through a joint XRT--BAT fit, found that the peak 
energy $E_{\rm peak}\sim4.5$ keV lies in the XRT energy range, and the
spectrum can be fitted by the sum of a black--body (BB) and a cut--off
power law (CPL).  
We show in Fig. \ref{060218} the X--ray data as
re--analysed by us, together with the best fit model (dashed lines).
The duration of this burst was so long that the BAT ``burst''
mode data contained only the first $\sim$ 350 sec of its emission. The
study of the following portion of the burst requires the analysis of
the ``survey'' mode data which are more difficult to handle, we
believe that the BAT team can analyse these data more appropriately
than us.
However, the BAT team already gives the total BAT fluence for this
burst (Campana et al. 2006).  We have then fitted the XRT time
integrated spectrum with the same model (i.e. BB+CPL and same 
$N_{\rm H}=6\times10^{21}$ cm$^{-2}$ -- see Campana et al. 2006), 
using the BAT fluence (Campana, priv. comm.; 
Dai, Zhang \& Liang 2006) as a constrain.
In this way we have obtained the same $E_{\rm peak}$ and 0.3--10 keV
fluence of Campana et al. (2006).

Through this ``BAT--fluence--constrained'' fit procedure we
analyzed the XRT data in the same time intervals reported in Campana
et al. (2006). 
Fig. \ref{060218} shows three spectra: the top one is the spectrum
integrated over the total observed duration of the bursts, while the
other two spectra correspond to the [159--309 s] and [2456--2748 s] time bins.  
These are
the first and the last spectra that can be obtained with both XRT and
BAT data, according to the time binning chosen by Campana et
al. (2006), which can be also seen in the right bottom panel of
Fig. \ref{060218}.  For the [159--309 s] time--bin the BAT ``burst" mode
data were available and we can then show the corresponding XRT+BAT
data.  For the [2456--2748 s] time--bin we only show the XRT data, 
fitted with the same model and procedure (i.e. forcing the fit 
to have the same BAT fluence reported in Campana et al. 2006) 
as for the time--integrated spectrum.

Note also that for all the three spectra (first and last time bin and
time integrated) we show the BB component resulting from the fit
(dotted lines).  This component contributes at least 20\% to the total
spectral fluence, and it has been interpreted as due to the
jet shock breakout (Campana et al. 2006). 
For this, we will not consider it in our spectral--evolution model.

In the right bottom panel of Fig. \ref{060218} we show the light curves
in the BAT and XRT energy ranges and also the evolution of $E_{\rm c}$
(i.e. $E_{\rm c}=E_{\rm peak}/(1-\alpha)$, where $\alpha$ is the energy
spectral index of the CPL model).
The values of $E_{\rm c}$ have been derived with the same
procedure discussed above.

To model the time evolution of this burst we need
the normalisation $K$ and the spectral index $\alpha$ 
of the power law, and
the cut--off energy $E_{\rm c}$, as a function of time.
For the latter we have simply interpolated the derived values
with a smooth function (the combination of a very weakly rising and 
a decaying power law of time: solid line in the right bottom panel 
of Fig. \ref{060218}), while the assumed behaviour of $K$ and $\alpha$
are shown in the left bottom panel of the same figure.
We have chosen functions which are either constants or power laws
of time.

Having reconstructed in this way the entire time evolution of this
burst, we could then sum--up the instantaneous spectra to
give the spectra integrated in specific time bins and the
spectrum integrated over the entire burst duration.
Note that the time integrated spectra {\it are not necessarily
described by a BB+CPL model, since the spectral parameters evolve.}
This can be the reason why, quite often, one obtains a better fit
with a time resolved spectrum rather than with a time integrated one.

The agreement with the observation is excellent, but expected, since
what we have done is merely to interpolate with smooth functions of
time the real evolution of the spectral parameters.  The
overprediction of the flux at high energies in the time integrated
spectrum (solid-red line in top panel) is due to the fact that the
smooth function interpolating $E_{\rm c}$ lies above the data points in the
second part of the time evolution (see the right bottom panel of
Fig. \ref{060218}).

\subsection{GRB 031203}

\begin{figure*}
\vskip -1 true cm
\centerline{\psfig{file=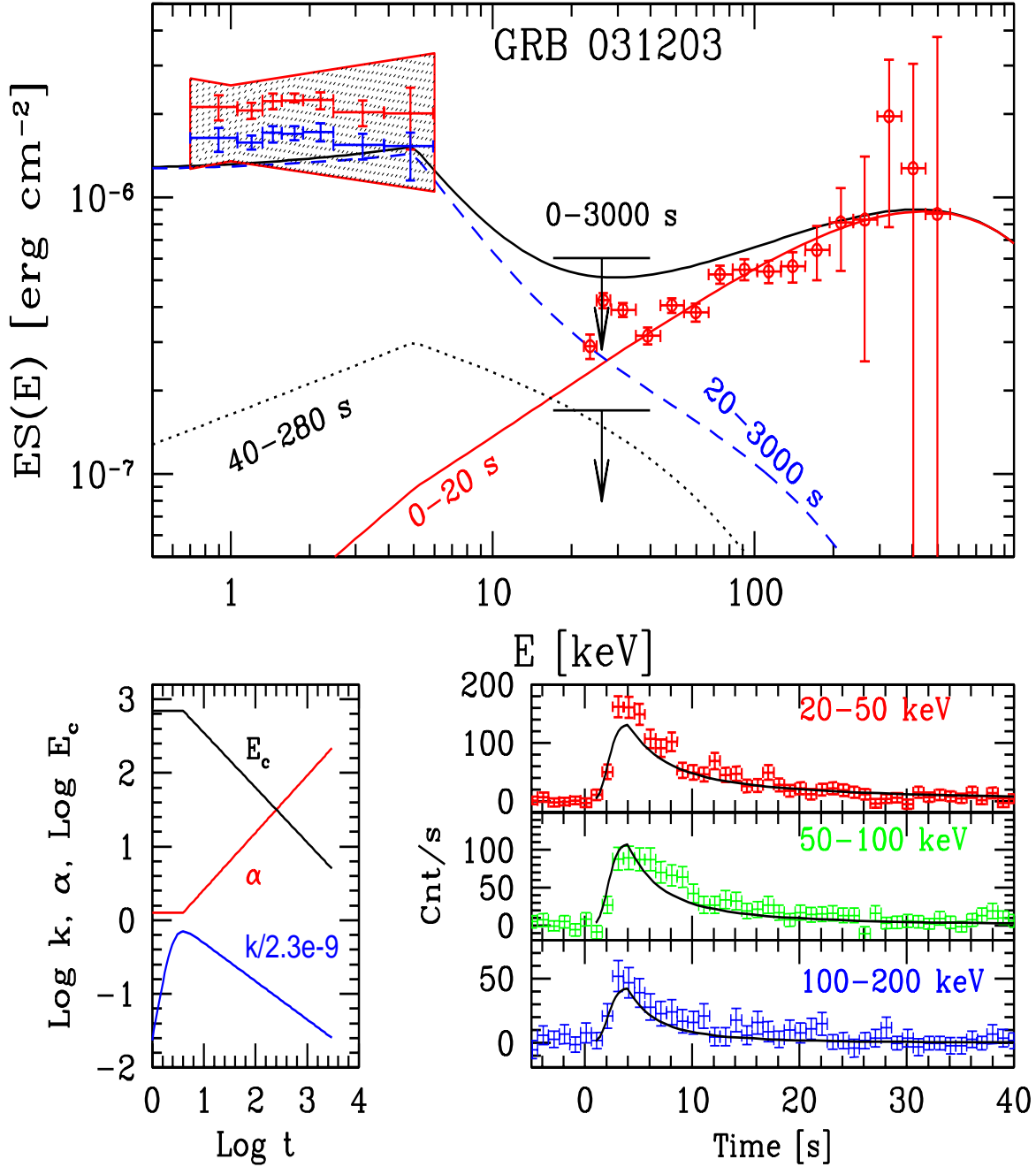,width=17truecm,height=21truecm}}
\vskip -1.5 true cm
\caption{
Top panel: the spectral energy distribution of GRB 031203 is compared with
the result of our model.
Here we show the INTEGRAL data as analysed by us
and the low energy emission as inferred by the light scattering echo
calculated by Watson et al. (2006).
The bow--tie corresponds to the uncertainty in the amount of the 
scattering material. 
In addition to those, there are additional uncertainties connected to
which cross section is used, as discussed in Tiengo \& Mereghetti (2006).
The lines corresponds to the emission time integrated in different 
energy intervals, as labelled. 
The bottom left panel shows how the normalisation $K(t)$, the 
high energy spectral index $\alpha$ and the cut--off energy $E_{\rm c}$ 
change in time.
The right bottom panel compares the observed and the calculated light 
curves in the three labelled energy bands. 
}
\label{031203}
\end{figure*}

This burst was discovered by INTEGRAL, and the spectrum in 
the 20--400 keV band is shown in Fig. \ref{031203}.
In this figure we show the spectrum as reanalysed by us
(see. Section 3.2).

The soft X--ray spectrum is inferred by the light scattering echo 
discovered by XMM--Newton (Watson et al. 2004) and modelled
by Watson et al. (2006) to get the intrinsic spectrum 
illuminating the galactic dust clouds.
Note that Tiengo \& Mereghetti (2006) re--derived the
illuminating spectrum and fluence by adopting a different
cross section for the scattering process between X--ray 
photons and dust, obtaining a fluence somewhat smaller than 
Watson et al. (2006), but with a similar spectrum.
Therefore both groups agrees that the soft (less than a few keV) 
X--ray emission lies much above the extrapolation from 
the high energy spectrum, and that the slope of the 
spectrum below a few keV is softer than the 20--200 keV slope.

The lightcurves (in counts per second) in tree different energy bands
have been extracted by us using the most recent version of the 
response matrix.

The modelling of the spectral evolution of this burst is aimed to reproduce:
i) the time integrated spectrum as seen by INTEGRAL in the 20--200 keV band;
ii) the upper limit in the medium energy band [17--40 keV];
iii) the soft X--ray emission as inferred by the light scattering echo and
iv) the light curves (in counts per second) in three different
energy bands of INTEGRAL [20--50 keV; 50--100 keV and 100--200 keV].

\subsubsection{Modelling the spectral evolution}

We model the spectral evolution of the prompt emission 
assuming that the spectrum, at any time, is a broken power law
ending with an exponential cut.
We assume a broken power law (instead of a single one) to
limit the total energetics to a finite value.
For simplicity, we assume that the energy spectral index
$\alpha$ of the low energy branch is equal to 1/2 the value
of the energy spectral index of the high energy part
(we do this just to limit the number of free parameters).
The two power laws connect at the break energy $E_{\rm b}$.
We assume that the cut--off energy $E_{\rm c}$, the break energy 
$E_{\rm b}$,  the power law index $\alpha$ and the normalisation 
can evolve in time. 
We set
\begin{eqnarray}
F(E, t) &=& K(t)
\left( {E \over E_{\rm b}}\right)^{-\alpha(t)/2};
\,\,\,\qquad\, E \le E_{\rm b}  \nonumber \\  
F(E, t) &=& K(t)
\left( {E \over E_{\rm b}} \right)^{-\alpha(t)} \times \nonumber \\ 
 &~& \exp\left[-{E-E_{\rm b} \over E_{\rm c}(t)}\right];
\qquad \qquad E \ge E_{\rm b}
\end{eqnarray}
where the normalisation $K$ is 
\begin{equation}
K(t)\, =\, {(t/t_m)^{b_1}\over 1+(t/t_m)^{b_1+b_2}} 
\end{equation}
here $b_1$ and $b_2$ are indices
describing the evolution of the normalisation of the spectrum,
and $t_{\rm m}$ is related to the time $t_{\rm max}$ of the maximum of 
$K(t)$ by
\begin{equation}
t_{\rm max} \, =\,t_{\rm m} \left( {b_1\over b_2}\right)^{1/(b_1+b_2)}
\end{equation}
The spectral index $\alpha$ and the cut--off energy, $E_{\rm c}$, 
are assumed to evolve as power laws of time (i.e.
$\alpha \propto t^a$, $E_{\rm c}\propto t^{-e}$,
but we allow for the possibility that they remain initially
constant (for a time equal or shorter than $t_{\rm max}$).
All these choices are arbitrary, the only guiding line
is simplicity.

The constant $K$ is chosen to match the observed spectral shape
of the fluence the burst obtained by the high energy data.

To reconstruct the light curves in one band $\Delta E= E_2 - E_1$ keV
in counts $s^{-1}$
we multiply $F(E, t)$ by the normalised effective area 
$A(E)$ of the INTEGRAL instruments to get
\begin{equation}
C(\Delta E, t) \, =\, \int_{E_1}^{E_2} { F(E,t) \over E} A(E) dE
\end{equation}
Then we re--normalise all the light curves obtained in different
energy bands by the same normalisation factor.

\subsubsection{Results for GRB 031203}

The left bottom panel of Fig. \ref{031203}
shows the behaviour of $K(t)$, $\alpha(t)$ and the cutoff energy 
$E_{\rm c}(t)$
which we have assumed for the spectral evolution of GRB 031203.

The top panel of Fig. \ref{031203} shows the SED resulting from our 
modelling, while the right bottom panel of the same figure reports
the light curves of the model (solid lines)
in different bands and compares those with the data.
The assumed overall duration of the prompt emission is 3000 seconds:
of those, only the first $\sim 50$ seconds have been detectable
by INTEGRAL in the 20--200 keV energy band.

We can see that the assumed hard--to--soft spectral evolution
is in agreement with all the information we have about the prompt 
emission of this burst, and predicts a peak energy of the
overall spectrum at $E_{\rm peak}\sim 5$ keV.
Also the total energetics is enhanced if we include the soft
X--ray prompt emission. 
This is shown in Fig. \ref{correla}, where one can see
how GRB 031203 ``moves" in the $E_{\rm iso}$--$E_{\rm peak}$ plane.
With the derived values of these parameters  
GRB 031203 obeys the Amati relation.

\begin{figure}
\vskip -0.5 true cm
\centerline{\psfig{file=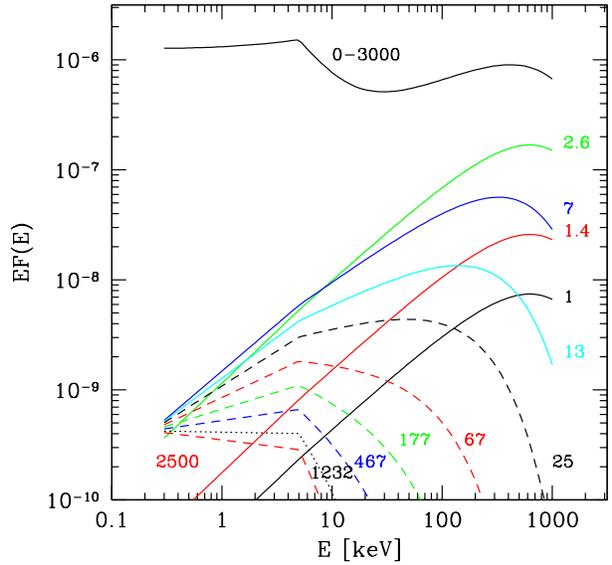,width=9truecm,height=9truecm}}
\vskip -0.7 true cm
\caption{
Calculated spectra of GRB 031203 at different times, 
as labelled (number are seconds). Also shown is the time integrated
spectrum (time interval 0--3000 seconds).
}
\label{spectra031203}
\end{figure}

\subsection{GRB 980425}

\begin{figure*}
\vskip -1 true cm
\centerline{\psfig{file=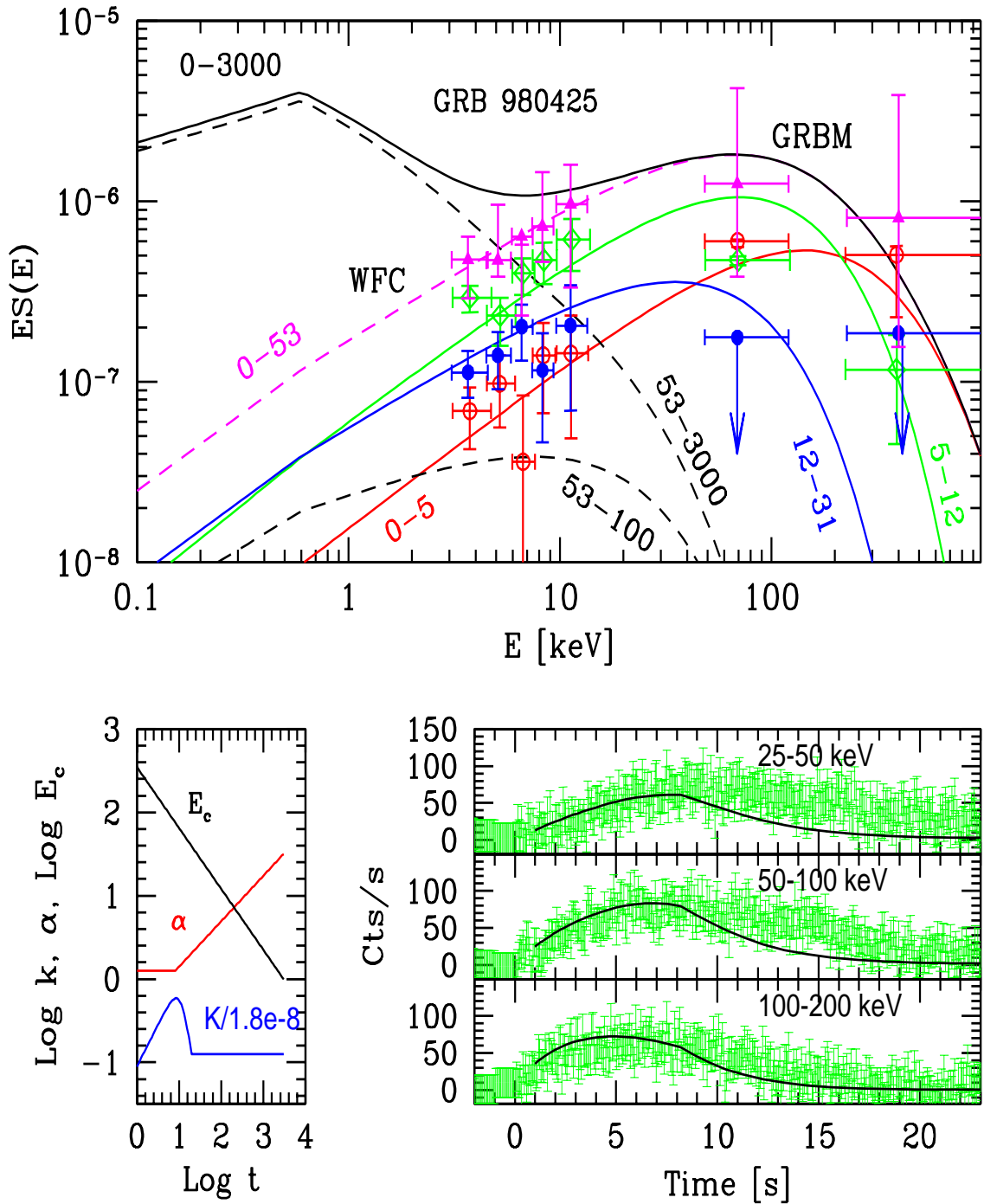,width=17truecm,height=21truecm}}
\vskip -1.5 true cm
\caption{
Top panel: the spectral energy distribution of GRB 980425 is
compared with the result of our model.  Here we show the $Beppo$SAX
data, as originally reported by Frontera et al. (2000), in three
different time--intervals, and the spectrum time--integrated over the
duration of the burst as seen by $Beppo$SAX.  The bottom left panel
shows how the normalisation $K(t)$, the high energy spectral index
$\alpha$ and the cut--off energy $E_{\rm c}$ change in time.  The right
bottom panel compares the observed and the calculated light curves in
the three labelled energy bands.  The observed light curves are from
BATSE.  }
\label{980425}
\end{figure*}

The prompt emission of this burst was detected both by BATSE and by
$Beppo$SAX. The latter detected it both with the Gamma Ray Burst
Monitor (GRBM) and the Wide Field Camera (WFC) (e.g. Pian et
al. 1999).  While in Fig. \ref{outliers} we show the BATSE spectrum
(as re--analysed by us), in the top panel of Fig. \ref{980425} we show
the $Beppo$SAX data (Frontera et al. 2000).  This is because we are
here interested in the medium energy X--ray band more than the high
energy one.  
The time integrated fluences of BATSE and the GRBM instrument are 
consistent.
We also show the spectra time--integrated in 3 different time intervals, 
using the data published in Frontera et al. (2000).

In the right bottom panel of Fig. \ref{980425} we show the light curve
of the count rate as detected by BATSE, in three different energy
bands [25--50 keV; 50--100 keV and 100--200 keV].

We have then applied our toy model using the same approach described
in Sec. 4.2.1.  The evolution of $K(t)$ $\alpha(t)$ and $E_{\rm c}(t)$ is
shown in the left bottom panel of Fig. \ref{980425}.

The resulting spectra are shown as solid lines in the top panel of
Fig. \ref{980425}.  They have been time integrated in the same time
intervals as the data.  We also show the spectrum corresponding to the
complete time evolution, assumed to last for 3000 seconds.  We have
chosen this duration just because it is equal to the one assumed for
GRB 031203 and observed for GRB 060218.  The calculated light curves
are overlapped to the BATSE 64ms observed light curves in the
right bottom panel of Fig. \ref{980425}.

We can see that, again, the calculated spectra and light curves are
consistent with what observed.  For this burst, unfortunately, we do
not have any supplementary information forcing us to believe that the
duration of the prompt emission is long (i.e. the WFC was not
sensitive enough to follow the burst for a long time, and there was no
information about late emission as was the case of the ``echo rings"
for GRB 031203).

On the other hand, our point is the following: if GRB 980425 is
similar to GRB 031203 (and both are similar to GRB 060218), then it is
possible that its peak energy $E_{\rm peak}$ is much smaller than what
derived by the high energy spectra only.  
In the case we have just shown $E_{\rm peak} \sim 0.6$ keV, 
once the spectrum is integrated over the entire (assumed) duration.  
Were this the case, then GRB
980425 would not be an outlier any longer, as can be seen in
Fig. \ref{correla}, where we show the new location of this burst.

\begin{table}
\begin{tabular}{llll}
\hline & GRB 980425 &GRB 031203 &\\ 
Observed & & & \\ 
$E_{\rm peak}$ &138 &291 &keV \\ 
$E_{\rm iso}$ &1.62$\times 10^{48}$ &1.2$\times 10^{49}$ & erg \\ 
\hline 
Scattering & & & \\ 
$E_{\rm peak}$ &25 &30 & keV\\ 
$E_{\rm iso}$ &9.35$\times 10^{50}$ &1.93$\times 10^{51}$ &erg\\
\hline 
Spectr. Evo.  & & & \\ 
$E_{\rm peak}$ &0.6 &5 &keV\\ 
$E_{\rm iso}$ &2.54$\times 10^{48}$ &1.6$\times 10^{50}$&erg \\ 
\hline
\end{tabular}
\caption{
The value of $E_{\rm peak}$ and $E_{\rm iso}$ that are 
observed and that are predicted by the scattering and spectral 
evolution models discussed in this paper. 
These are the values used in Fig \ref{correla}.
Note that the $E_{\rm iso}$ (observed) for GRB 031203 takes into account
the small reduction due to the re-analysis of the INTEGRAL data.
}
\end{table}

\section{Discussion}

In this work we have explored three different scenarios 
to see if GRB 980425 and GRB 031203 are really outliers 
with respect to the Amati relation (and therefore also
outliers to the Ghirlanda relation) or if they only
{\it appear} as such.

The first model (off axis viewing angle) can bring the two bursts on
the Amati relation only if the energetics of these bursts (as would be
seen by an on--axis observer) is very large.  
Such bursts are not
expected to exist at low redshifts, and then this argues against this
model
{as an explanation for the two bursts being only ``apparent" outliers.
}

Then we explored the idea that these bursts have ``normal" energetics
and values of $E_{\rm peak}$, but both these quantities are modified
by a scattering cloud located in the vicinity of the bursts.  It must
be located in the vicinity of the burst because in this case the
prompt emission can completely ionise the material (Lazzati \& Perna
2002), making the soft X--ray afterglow insensitive to absorption.  
If very close to the burst, the scattering material would be accelerated
to large velocities. 
In the comoving frame of the moving material, incoming photons
would be seen redshifted, and a larger fraction of photons would
scatter in the Thomson regime.
The transmitted spectrum would then be similar (albeit not
identical) to the one obtained in the case of a not moving 
material with a larger $\tau_{\rm T}$.
The complete and exact analysis implies to assume the
location and width of the scattering screen, which would
determine its acceleration, and this lies outside the aim
of this paper.

In this model the intrinsic (i.e. before scattering) energetics is
greater than the observed one, but not by a large factor.  
Even with a substantial value of the scattering optical depth 
(i.e. $\tau_{\rm T}\sim$a few),
the reduction in flux of the transmitted radiation is of the order of
one--a few hundreds.  
The scattering process is more effective on low
energy X--ray photons, implying that the transmitted spectrum is
``bluer" than the intrinsic one.  
Therefore the intrinsic $E_{\rm peak}$ can be smaller than observed.  
The two things together
(i.e. intrinsically, the two bursts are more energetic and much
redder) make it possible for them to lie on the Amati relation. 
The increase in energetics is not so large to face 
the problem of having too energetic bursts at small redshifts.

The problem, for this idea, is to explain why these bursts are surrounded
by dense and thick material (for scattering), while other bursts are not.
The partial answer we can give is to consider one obvious selection 
effect: we see distant bursts only if they are surrounded by material with a 
small value of $\tau_{\rm T}$, since otherwise they go under 
the detection threshold.
Consider that even values of $\tau_{\rm T}\sim 1$ have almost no effect
on the observed spectrum, since in this case more than half of the photons 
pass through the scattering cloud undisturbed (see Fig. 3).
Even the fast variability, if present, would not be smeared out,
since the received flux is dominated by the transmitted photons:
the scattered ones, having a longer path to travel, arrive 
later and more diluted in time (i.e. with a much reduced flux).

On the other hand, this model poses several other problems.
The most severe of them is that we expect some signature
of the presence of such a thick screen in the vicinity
of the burst, produced by the collision of the bursts and
supernova ejecta with the screen itself.

The third model we have proposed is the most promising, especially
after GRB 060218, which we used as a guide to model the two outliers.
In this framework the derived intrinsic powers and energetics are the
least demanding, as can be seen in Fig. \ref{correla} and in Table 2,
where we list the values of $E_{\rm iso}$ and $E_{\rm peak}$ for the
two bursts.  
In the framework of this model the presence of both a
hard and a soft component (in GRB 031203) is the result of the
spectrum evolving in time, and is not due to two separated components
(i.e. two emission mechanisms).  Note that the real bolometric and
time integrated energetics (i.e. $E_{\rm iso}$) is larger than what
derived by considering the high energy spectrum only, but not by a
large factor.

The main result of our study is that GRB 031203 and GRB 980425 are
likely to be {\it apparent} outliers: their intrinsic properties are
instead consistent with the Amati relation (and therefore it is
possible that they obey even the Ghirlanda relation, and they would do
so with jet opening angles not particularly extreme, see
Fig. \ref{correla}, although larger than the average values found
for ``normal'' GRBs).  This possibility is strengthened by
the fact that GRB 060218, energetically a twin of GRB 031203, lies on
the Amati relation independently of our modelling.  
GRB 980425 and GRB 031203 could also become consistent with the
newly found correlation between the peak luminosity of the prompt, 
$E_{\rm peak}$ and the ``high signal" timescale $T_{0.45}$ 
(Firmani et al. 2006),
if they indeed have $E_{\rm peak}$ around $\sim$1 keV.

On the other hand these two bursts, together with GRB 060218, 
besides being underluminous, have a duration much longer than the 
other bursts, including the X--ray rich and the X--ray flash 
population of bursts (see e.g. Lamb, Donaghy \& Graziani 2005).
The inevitable conclusion is that the Amati relation (and possibly the
Ghirlanda one) is more robust than previously thought, being obeyed by
bursts that are underluminous and with a duration longer than the
average one and that (with respect to the Amati relation)
GRB 980425 and GRB 031203 are not
representative of a different population of objects with respect to
``classical'' GRBs.

\section*{acknowledgements}
We thank the Italian INAF for financial support.

\end{document}